\documentclass[floatfix,%
 reprint,
 amsmath,amssymb,
 aps,
]{revtex4-2}

\usepackage{graphicx}
\usepackage{dcolumn}
\usepackage{bm}
\usepackage{algorithm}
\usepackage{algpseudocode}
\usepackage{appendix}
\usepackage{xcolor}

\usepackage{mathtools}

\begin{document}

\preprint{APS/123-QED}

\title{Active lattice percolation: an apparently new universality class of percolation transitions}

\author{Vernon M. Hughes}
\author{David A. Huse}

\affiliation{Department of Physics, Princeton University, Princeton, NJ, 08544, USA}

\date{\today}

\begin{abstract}
We investigate a class of dynamic percolation models on a lattice, in which connections are added and removed by a set of local stochastic rules.  In particular, the removal of any connection that can result in one connected cluster becoming two (or more) separate clusters is strictly forbidden.  These models include and are inspired by a model for local quantum error correction for a toric code studied by Chirame, \textit{et al.}, PRX
Quantum {\bf 6}, 030363 (2025), who identified a phase transition with first-order characteristics, such as bistability and a discontinuous order parameter. We show numerically that this transition is a hybrid percolation transition (HPT), having characteristics of a continuous percolation transition such as diverging critical clusters while at the same transition having discontinuities like a first-order transition.  We then modify this model to be fully isotropic, showing that the HPT remains.  In both of these models, the steady state on one side of the HPT is an absorbing state; by turning on additional isotropic local stochastic moves, we remove the absorbing state and strongly suppress or remove the first-order transition, leaving an apparently continuous percolation transition. 
Curiously, the numerically-observed percolation critical exponents do not change significantly between the models with hybrid transitions and the model with an apparently continuous transition.  These exponents are significantly different from known percolation transitions, indicating that this may be a new universality class. 
\end{abstract}

\maketitle

\section{Introduction}

Non-equilibrium phase transitions arise in dynamical systems across condensed matter and statistical physics \cite{broadbent1957percolation,hinrichsen2000nonequilibrium,odor2004universality,takeuchi2007directed}, biology \cite{fang2019nonequilibrium}, and fault-tolerant computation \cite{dennis2002topological}. One direction of interest in such non-equilibrium systems is whether distinct phases can be stabilized in the presence of noise or perturbations via local classical dynamics. In the context of fault-tolerant computation, such systems represent robust memories or autonomous local error-correcting codes \cite{toom1980stable,gacs1986reliable,chirame2025stabilizing,lake2026squeezing}. Often defined as probabilistic cellular automata, these systems have been studied through the lens of asynchronous updates, describing continuous-time Markov processes, and have revealed different continuous or first-order phase transitions \cite{hinrichsen2000nonequilibrium,odor2004universality,lake2026squeezing,guo2007gtpc}.

One such model, motivated by local quantum error correction in the presence of erasure noise, was recently proposed by Chirame \textit{et al.} \cite{chirame2025stabilizing}. For a toric code on a honeycomb lattice, this model contains a phase with stabilized logical information, and a phase transition to an absorbing state where logical information is lost. This phase transition was numerically found to be first-order, with a strong discontinuity in its steady-state properties  \cite{chirame2025stabilizing}. However, no specific instability was identified, leaving open questions as to whether the first-order transition is driven by nucleation or by some other mechanism, and whether this transition may also show some critical behavior similar to continuous phase transitions.

Models with a first-order-like discontinuous order parameter can also exhibit critical behavior via a diverging correlation length.  For percolation transitions, such behaviors have been labeled hybrid percolation transitions (HPTs) \cite{cho2026recent}. These hybrid transitions have typically been found in mean-field and network-based models, such as $k$-core and bootstrap percolation \cite{lee2017universalHPT,chalupa1979bootstrap}, explosive percolation \cite{panagiotou2011explosive,cho2016hybrid}, generalized epidemic processes \cite{janssen2004generalized,lee2017universalHPT,grassberger2016phase}, neural networks \cite{goltsev2010stochastic}, and cascading failures on interdependent networks \cite{buldyrev2010catastrophic,son2011percolation,baxter2012avalanche}. Some of these models have finite-dimensional variants with HPTs in $d\ge 3$ dimensions, or in $d=2$ dimensions (2D) with long-ranged contacts \cite{grassberger2016phase,lee2016hybrid}. Other finite-dimensional models on regular lattices include jamming percolation and kinetically constrained models in 2D and 3D, where divergence is faster than a power law, thereby not admitting critical exponents \cite{toninelli2006jamming,toninelli2008new,biroli2008spiral,ghosh2014jamming}, and no-enclaves percolation, a static model where the area within any enclosed cluster starting from ordinary percolation is filled in (2D) \cite{sheinman2015gels,hu2016nep}. We are not aware of a 2D lattice model with local non-equilibrium dynamics exhibiting a HPT similar to the transitions that we study here.

In this work, we study a class of models inspired by Chirame \textit{et al.}~\cite{chirame2025stabilizing}, where the local non-equilibrium dynamics can add, remove, or re-route connections.  Removal of a local connection that could disconnect a cluster into two (or more) different connected clusters is forbidden.  In Sec.~\ref{sec:nw}, we show that the local erasure flag model from Chirame \textit{et al.} has critical characteristics, with a diverging length scale at the phase transition. Therefore, we propose that its phase transition is actually a hybrid percolation transition. This original model has strong spatial anisotropy inspired by Toom's rule \cite{toom1980stable}, and we show in Sec.~\ref{sec:six_fold} that adapting the model to be spatially isotropic keeps the HPT with no numerically detectable change in the critical exponents.  Finally, in Sec.~\ref{sec:remove_absorb}, we remove the absorbing state from this isotropic model by allowing for additional isotropic local stochastic moves, finding that the strong first-order nature of the transition goes away but that the model still features critical behavior with critical exponents that remain the same to within our numerical precision.  Since these critical exponents do not match previously-known  percolation transitions, these phase transitions appear to be in a new universality class.  We suspect that the key ingredient that produces this apparently new universality class is the local stochastic dynamics that is forbidden to break up connected clusters.  But this is after exploring only three variations on this model, and explorations in other directions may still be needed to elucidate what are the important ingredients of this type of model that set the universality class(es) of these phase transitions.

\section{Toom-inspired Model}
\label{sec:nw}

We work on a honeycomb lattice with unit length edges. We define vectors $\vec{a}$ and $\vec{b}$, which are vectors of length three that are aligned with the underlying lattice and separated by an angle of $\pi/3$ (see Fig.~\ref{fig:fig1}(a)).  We use periodic boundary conditions such that the positions $\vec{r}$, $\vec{r}+(L\vec{a}/3)$, and $\vec{r}+(L\vec{b}/3)$ are all the same position, with $L$ an integer multiple of 3.
This lattice has $L^2$ edges (of unit length), $2L^2/3$ vertices, and $L^2/3$ plaquettes. The honeycomb lattice can also be visualized as two triangular sublattices, represented by purple and yellow vertices in Fig.~\ref{fig:fig1}(a).

\subsection{Model}

In the toric code model from Chirame \textit{et al.}~\cite{chirame2025stabilizing}, physical qubits lie on the edges of this lattice, with stabilizers at vertices or plaquettes. In addition to the stabilizers, erasure flags are used in order to perform local error correction and stabilize a continuous depolarizing noise channel. Each erasure flag, which takes on a classical binary value, flags a physical qubit, so these flags also live on the lattice edges.  When this is viewed as a percolation model, it is bond percolation, with the flagged bonds being occupied.

\begin{figure}[!t]
    \centering
    \includegraphics[width=\columnwidth]{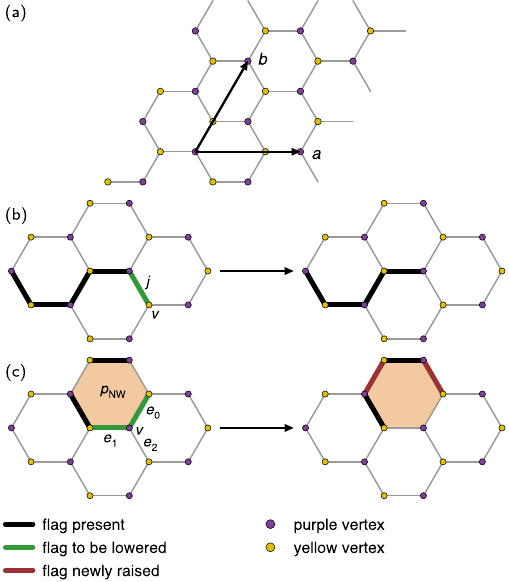}
    \caption{Lattice geometry and flag lowering moves for the Toom-inspired model: (a) A patch of the honeycomb lattice on a torus with $L=6$. The vectors $\vec{a}$ and $\vec{b}$ define the periodic boundary conditions, as discussed in the main text. The purple and yellow vertices represent two underlying triangular lattices. (b) \textit{Leaf} correction move: For a chosen vertex, $v$, if only a single flag is raised among the three edges connected to $v$, that flag is then lowered (edge $j$). (c) \textit{Loop} correction move: The loop correction move is only applied at purple vertices. If $v$ is a purple vertex, there is a plaquette to the northwest of $v$, denoted $p_{\text{NW}}$. We label the edges connected to $v$ as shown, where $e_2$ does not border $p_{\text{NW}}$. If $e_0$ and $e_1$ have flags raised and $e_2$ does not, then the flags on $e_0$ and $e_1$ are lowered, and flags are raised on all the other edges bordering $p_{\text{NW}}$ (if they were not already).}
    \label{fig:fig1}
\end{figure}

The noise and correction channels in \cite{chirame2025stabilizing} are both modeled as continuous-time Lindbladians, where each channel acts on both the erasure flags and stabilizers. The flag dynamics alone are sufficient for identifying the phase transition \cite{chirame2025stabilizing}, hence we will only describe the actions of each channel on the flags.  This results in a classical stochastic lattice model, and the quantum aspects of the original motivating model may be ignored. The noise channel simply raises a flag on an edge, $j$, provided it was not already flagged, at a rate $\eta=1$.  So connections are added on each bond at this rate. There are two types of correction moves, neither of which can break connectivity of clusters of flags on the lattice, in order to be compatible with tracking stabilizers. The correction moves are applied after selecting a vertex, $v$, and considering the three edges connected to $v$. If only one edge, $j$, is flagged, then that flag can be lowered; this is termed a \textit{leaf} move (Fig.~\ref{fig:fig1}(b)). The leaf move alone is insufficient for stabilizing logical information, since it cannot remove flags on any closed loops. Thus, a second type of correction move, known as a \textit{loop} move, is introduced to act on vertices $v$ that are connected to exactly two flagged edges. The loop move only acts on vertices in the purple triangular sublattice and only acts when the two flagged edges border the plaquette to the northwest (NW) of $v$ (see Fig.~\ref{fig:fig1}(c)). Then, the flags on both of these edges can be lowered, but flags on all of the other edges bordering the northwest plaquette must be raised if they were not already. This ensures that connectivity is never broken; it is instead re-routed. The anisotropy between the purple and yellow sublattices, and the loop move's NW directionality was introduced to mimic Toom's rule~\cite{chirame2025stabilizing,toom1980stable}; as a consequence of this directionality, connected clusters on average move towards the NW with some drift speed, while they also fluctuate in shape and size due to the noise and correction moves.  The correction moves are attempted at a per-edge rate of $\gamma$.

\subsection{First-order nature}

\begin{figure}[!t]
    \centering
    \includegraphics[width=\columnwidth]{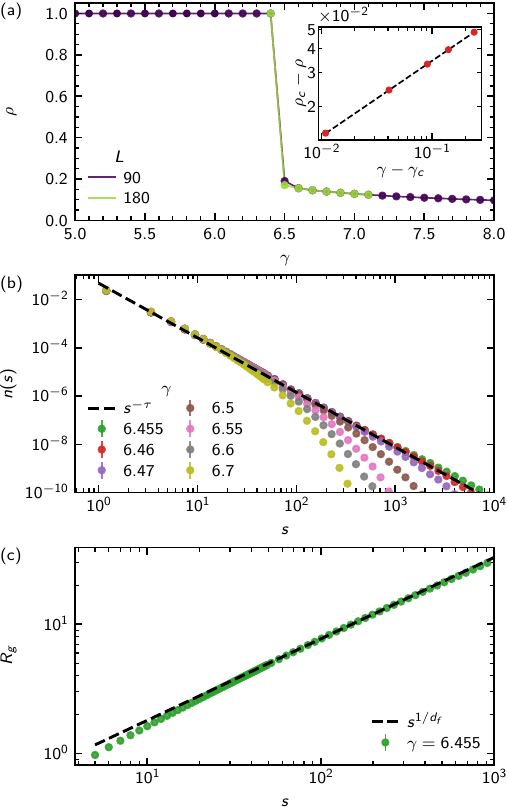}
    \caption{Toom-inspired model data: (a) Steady-state flag density for different $\gamma$ at $L=90$ and $L=180$. Each point is averaged over 100 Monte Carlo trials at $t=3200$ (chosen such that $\rho(\gamma=6.4)=1$ for all 100 trials at $\gamma=6.4$ for each $L$).  Initial states had $\rho=0$. Inset: the average flag density for $L=750$ at $t=2000$. The black dashed curve is a fit of the form $\rho_c-\rho=A(\gamma-\gamma_c)^{1-\alpha}$ with $\alpha=0.61(5)$, $\rho_c=0.195(3)$, and $\gamma_c=6.459(3)$ ($\gamma_c$ determined independently from Fig.~\ref{fig:fig3}). (b) Determination of the Fisher exponent, $\tau$, near the transition at $\gamma_c$. The number of edges in each cluster is denoted as $s$, while $n(s)=\frac{N(s)}{L^2}$ where $N(s)$ is the number of $s$-clusters and $L^2$ is the number of edges on the lattice. Different colors correspond to different $\gamma$ values, and all points are averaged as described in Appendix~\ref{sec:appendix_cluster_analysis} over 100 Monte Carlo trials at $L=750$ with $t\in[1000,2000]$. The black dashed line is a fit to the $\gamma=6.455$ data of the form $n(s)=As^{-\tau}$, with $\tau=2.26(10)$. (c) Determination of the fractal dimension, $d_f$, near the critical point. For each cluster, $s$ is the number of edges in that cluster and $R_g$ is its radius of gyration. Each data point is averaged over 100 Monte Carlo trials at $L=750$ for $t\in[1000,2000]$. The black dashed line is a fit of the form $\log  R_g =\frac{1}{d_f}\log s+B$, yielding $d_f=1.59(8)$.}
    \label{fig:fig2}
\end{figure}

Using Monte Carlo simulations, Chirame, \textit{et al.} identified a phase transition between an absorbing-state phase, where every flag is raised in the steady state and thus no correction moves can be applied, and an active phase where the average flag density remains low. The active phase is present for $\gamma>\gamma_c$ (with $\eta=1$) \cite{chirame2025stabilizing}. The authors numerically demonstrated that this is a first-order phase transition, where the steady-state flag density, $\rho$, is discontinuous, with two primary pieces of evidence. First, the time $T$ required for the average erasure flag density, $\rho$, to cross a certain large threshold after initializing with an empty lattice 
was found to be power-law in the system size, $L$, for $\gamma_c=6.45(2)$, sub-power-law for $\gamma<\gamma_c$, and exponential in $L$ for $\gamma>\gamma_c$. This implies that when taking the limit of the system size to infinity and approaching the steady-state by taking $t$ to be a power-law of $L$, the system will approach the absorbing state and a flag density of $1$ for $\gamma<\gamma_c$, while for $\gamma>\gamma_c$ the average flag density remains near or below $\rho=0.2$. This was further supported by bimodal histograms of the flag density near $\gamma_c$, with clear peaks near $\rho=0.2$ and at $\rho=1$ across a range of $L$ and time $t$.  With increasing $t$, the peak at lower flag density becomes depleted as trials end up in the absorbing state, suggesting a first-order transition with a discontinuous flag density in the steady state. The typical flag densities of the two phases can be seen in Fig.~\ref{fig:fig2}(a).

Using a large system size of $L=750$ initialized with $\rho=0$, we evaluate the steady-state $\rho(\gamma)$ near the transition and find that the derivative $\frac{d\rho}{d\gamma}$ appears to be diverging.  We fit the average flag density as $\rho=\rho_c-A(\gamma-\gamma_c)^{1-\alpha}$ and extract $\alpha=0.61(5)$.  We call this exponent $\alpha$ because we are making a loose analogy to standard (e.g. Ising model) critical points, with $\gamma$ the analog of temperature and $\rho$ the analog of the energy density.  It appears that for $\gamma<\gamma_c$, the flag density then jumps discontinuously from $\rho_c\approx 0.2$ to $\rho(\gamma<\gamma_c)=1$ (unlike a standard critical point), and that this $\gamma_c$ coincides with a percolation transition, as seen in the following section.
In HPT literature, the exponent describing the relationship between $\rho$ and $\gamma$ is typically denoted as $\beta=1-\alpha$ \cite{lee2017universalHPT}.

\subsection{Critical behavior}

In addition to this apparently continuous divergence of $d\rho/d\gamma$, we identify other continuous critical characteristics of this phase transition by looking at power-law scalings of connected flag clusters. These suggest that the phase transition is actually a hybrid-order percolation transition, with a discontinuous order parameter accompanied by criticality when the transition is approached from the active phase. In particular, we use large system sizes of $L=750$ initialized with $\rho=0$ in order to mitigate finite-size effects and prevent clusters from wrapping around the periodic boundary conditions of the lattice within reasonable timescales. This allows us to identify power-law scalings of fractal clusters, qualitatively similar to standard Bernoulli percolation clusters, but with quantitatively different critical exponents.

We study the typical mass and length scales of connected clusters near the transition using Monte Carlo simulations (details in Appendix~\ref{sec:appendix_monte_carlo} and~\ref{sec:appendix_cluster_analysis}). The mass of a cluster, $s$, is defined as the total number of edges belonging to that cluster, and the probability of having a cluster of a certain mass becomes power-law distributed at $\gamma_c$. We define the density of clusters of mass $s$ as $n(s)=\frac{N(s)}{L^2}$ where $N(s)$ is the number of $s$-clusters (at a given time) and $L^2$ is the number of lattice edges. The Fisher exponent, $\tau$, describes the power-law dependence of $n(s)=As^{-\tau}$ at $\gamma=\gamma_c$. We determine $\tau=2.26(10)$ from a fit to data at $\gamma=6.455$ and $L=750$ averaged over $t\in[1000,2000]$ and 100 Monte Carlo trials (Fig.~\ref{fig:fig2}(b)). The Fisher exponent for Bernoulli percolation in 2D is $\tau=\frac{187}{91}\approx2.05$ \cite{ahorony2003percolation}. For $\gamma>\gamma_c$, $n(s)$ falls off as an exponential in $s$ at large $s$, which is also consistent with percolation universality classes.  Further details on the convergence of $n(s)$ with respect to system size and simulation timescale can be found in Appendix~\ref{sec:appendix_convergence}.

\begin{figure}[!t]
    \centering
    \includegraphics[width=\columnwidth]{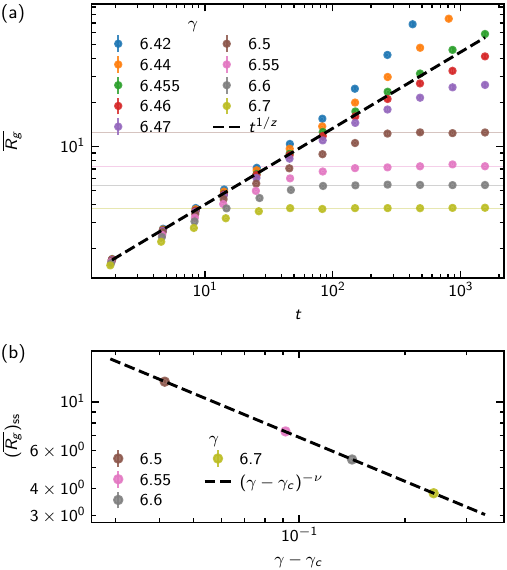}
    \caption{More Toom-inspired model data:  (a) The dynamic critical exponent, $z$. The mean radius of gyration over all clusters is defined as $\overline{R_g}=\sum_j (R_g)_j s_j^2/\sum_j s_j^2$ and computed for each cluster present at different times, $t$, over 100 Monte Carlo trials at $L=750$. Each data point corresponds to an average over a logarithmic time bin. The critical rate $\gamma_c$ is characterized by a power-law fit $\overline{R_g}=At^{1/z}$ at $\gamma=6.455$ with $z=1.91(6)$. The horizontal lines correspond to the steady-state mean radius of gyration for certain $\gamma>\gamma_c$, computed by averaging $\overline{R_g}$ over $t\in [1000,2000]$. (b) The steady-state mean radius of gyration, $(\overline{R_g})_{\text{ss}}$, as computed in (a) for $\gamma\ge 6.5$. The black dashed line is a fit of the form $(\overline{R_g})_{ss}=A(\gamma-\gamma_c)^{-\nu}$, yielding $\gamma_c=6.459(3)$ and $\nu=0.67(2)$.}
    \label{fig:fig3}
\end{figure}

We study the relationship between mass and length scales of clusters by looking at their fractal dimension, $d_f$. We define the radius of gyration of a cluster as
\begin{equation}
R_g=\sqrt{\sum_{k}\frac{(x_k-\overline{x})^2+(y_k-\overline{y})^2}{s}}~,
\end{equation}
where $(\overline{x},\overline{y})$ are the Cartesian coordinates of the center of mass of the cluster, $(x_k,y_k)$ are the coordinates of each edge in the cluster, and $s$ is the total number of edges in the cluster. We use a breadth-first search (BFS), which identifies all clusters as connected components, and reconstructs each connected component by `unwrapping' the periodic boundary conditions in order to get a representation of the cluster in coordinates where every edge is connected (Appendix~\ref{sec:appendix_cluster_analysis}). For clusters that wrap around the periodic boundary conditions of the lattice, $R_g$ is undefined, so we ensure that the lattice size is large enough so that clusters do not wrap around the lattice within the simulation timescale. The fractal dimension relates the mass and length measures, with $R_g\sim s^{1/d_f}$ in the limit $s\to\infty$. We determine a fractal dimension of $d_f=1.59(8)$, as depicted in Fig.~\ref{fig:fig2}(c), using data at $\gamma=6.455$ and $L=750$, averaging over 100 Monte Carlo trials within $t\in[1000,2000]$. Further details of this estimation can be found in Appendix~\ref{sec:appendix_convergence}. The fractal dimension of Bernoulli percolation in 2D is $d_f=\frac{91}{48}\approx1.896$ \cite{ahorony2003percolation}. This difference in the fractal dimension leads to qualitatively distinct cluster geometries between this model and Bernoulli percolation, with examples of large clusters shown in Appendix~\ref{sec:appendix_convergence}.

The single-length hypothesis states that there is only one relevant lengthscale in percolation-type universality classes which dominates the critical behavior \cite{ahorony2003percolation}.  Thus if we consider the critical distribution of clusters and consider all the clusters with, say, $\ell/2\leq R_g\leq\ell$, these clusters are assumed to have a density (counting the clusters, not the links in the clusters) proportional to $\ell^{-2}$.
This assumption results in the hyperscaling relation
\begin{equation}
\tau=1+\frac{d}{d_f}~.\label{eq:hyperscaling}
\end{equation}
We treated $\tau$ and $d_f$ as independent exponents, rather than assuming that hyperscaling holds as it does for ordinary percolation. Our results are consistent with hyperscaling, since for $d_f=1.59(8)$, Eq.~\ref{eq:hyperscaling} yields $\tau=2.26(6)$, matching the measured value of $\tau=2.26(10)$.

Since this is a non-equilibrium model, we can also analyze how critical clusters grow over time with a dynamic critical exponent, $z$. This relates length and time measures via $\xi \sim t^{1/z}$, where $\xi$ is the correlation length. To obtain a measure of the characteristic length, we use
\begin{equation}
\overline{R_g}(t)=\frac{\sum_{k}(R_g)_{k}s_k^2}{\sum_{k}s_k^2}~,\label{eq:rgbar}
\end{equation}
where the sum is taken over all clusters $k$ in a given trial at a given time. In Eq.~\ref{eq:rgbar}, the numerator and denominator both diverge at the critical point since $\sum_k s_k^2=\sum_s s^2 N(s)\approx L^2 \int_s s^{2-\tau}ds$, which diverges since $\tau<3$. Therefore, the expression is dominated by the largest clusters, which are on the order of the correlation length, $\xi$. For $\gamma<\gamma_c$, $\overline{R_g}(t)$ grows faster than a power of $t$, as the lattice fills in and the system ends up in the absorbing state. At $\gamma_c=6.455(15)$, $\overline{R_g}(t)$ grows as a power law in time, with dynamic exponent $z=1.91(6)$. Finally, for $\gamma>\gamma_c$, there is a finite correlation length, as $\overline{R_g}$ saturates with time. All data are averaged over 100 Monte Carlo trials at $L=750$.

\begin{figure}[!t]
    \centering
    \includegraphics[width=\columnwidth]{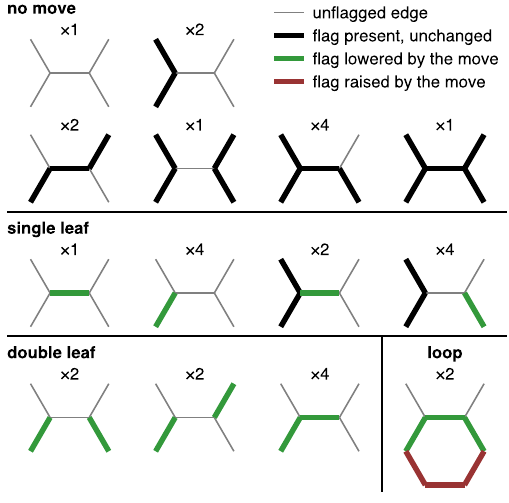}
    \caption{Flag lowering moves for the isotropic model. For a given edge, $j$, all 5 edges connected to the two vertices at the endpoints of $j$ are considered. This yields $2^5=32$ possible flag configurations, represented above with horizontal and vertical reflection symmetry. There are three types of moves, which do not break lattice connections and are guaranteed to never raise more flags than they lower. (i) \textit{Single leaf} moves lower one flag that is connected to a vertex where it is the only raised flag. (ii) \textit{Double leaf} moves lower two flags by applying leaf moves. (iii) \textit{Loop} moves are only applied in a configuration where three flags can be lowered, and at most three flags are raised. These loop moves are equivalent to the loop moves in the original model along with an extra leaf move, and are applied irrespective of the purple and yellow triangular sublattices.}
    \label{fig:fig4}
\end{figure}

We note that this diverging correlation length is defined in terms of clusters that are spatially connected at a given fixed time  in the limit of large system sizes and timescales.  This is to be contrasted with directed percolation, where clusters percolate in time: some population at directed percolation transitions survives for arbitrarily long times, but this population remains spatially sparse and disconnected at any fixed time, rather than forming a spatially connected percolation cluster. Indeed, previous work on the contact process has shown that in two spatial dimensions, the DP transition and an isotropic, geometric percolation point do not coincide~\cite{martin2020intermittent}. Therefore, the way in which we have defined $\tau$, $d_f$, and $z$ as geometric percolation exponents for geometrically connected clusters does not have direct analogues in directed percolation.

The finite correlation length for $\gamma>\gamma_c$ in percolation-type models allows us to classify the correlation length exponent, $\nu$. Indeed, for $\gamma\ge 6.5$, $\overline{R_g}$ saturates with time for $t\ge 1000$, and the saturated value $(\overline{R_g})_{\text{ss}}$ should vary as $(\gamma-\gamma_c)^{-\nu}$. We verify this in Fig.~\ref{fig:fig3}(b), where a joint fit to $\gamma_c$ and $\nu$ yields $\gamma_c=6.459(3)$ and $\nu=0.67(2)$. We employ this definition of $\nu$ since we can obtain a more precise estimate of $\gamma_c$ than $\rho_c$, as our fits for $\tau$ and $z$ help identify $\gamma_c$. However, given the earlier observed behavior that $\rho-\rho_c\sim (\gamma-\gamma_c)^{1-\alpha}$ with $\alpha=0.61(5)$, one could also define the correlation length exponent as $\xi\sim(\rho_c-\rho)^{\nu/(1-\alpha)}$ with $\nu/(1-\alpha)=1.7(2)$. This provides an alternative comparison to other percolation models, where the correlation length divergence is quantified in terms of $(p-p_c)$, and the correlation length exponent is $\nu=\frac{4}{3}$ for ordinary percolation and $\nu=0.73$ for directed percolation (although this exponent for directed percolation does not describe geometrically connected clusters, as discussed above) \cite{ahorony2003percolation,henkel2008nonequilibrium}.  

\section{An isotropic model}
\label{sec:six_fold}

\begin{figure}[!t]
    \centering
    \includegraphics[width=\columnwidth]{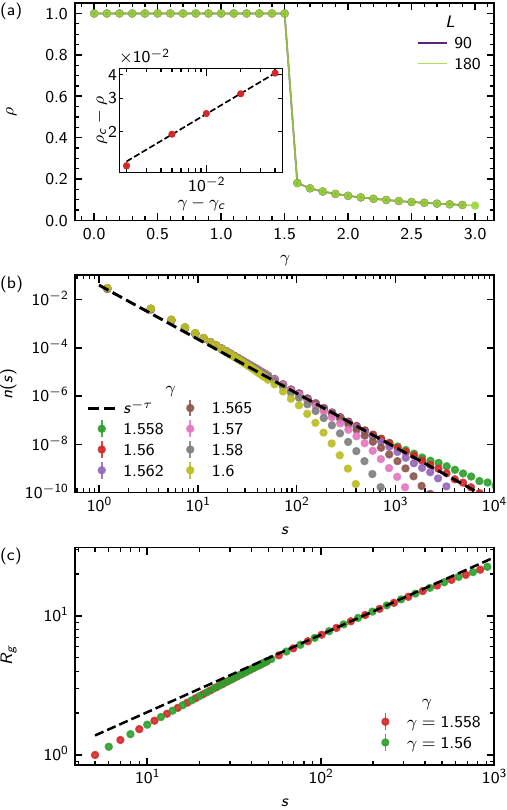}
    \caption{Isotropic model data: (a) Average flag density for different $\gamma$ at $L=90$ and $L=180$, at $t=500$. Each point is averaged over 100 Monte Carlo trials. Inset: the average flag density for $L=750$ at $t=2000$. The black dashed curve is a fit of the form $\rho_c-\rho=A(\gamma-\gamma_c)^{1-\alpha}$, with $\alpha=0.64(10)$, $\rho_c=0.221(12)$, and $\gamma_c=1.560(2)$ ($\gamma_c$ determined independently from Fig.~\ref{fig:fig6}). (b) Determination of the Fisher exponent, $\tau$, near the transition at $\gamma_c$. $s$ is the number of edges in each cluster, while $n(s)=\frac{N(s)}{L^2}$ where $N(s)$ is the number of clusters of size $s$ and $L^2$ is the number of edges on the lattice. Different colors correspond to different $\gamma$ values and all points are averaged as described in Appendix~\ref{sec:appendix_cluster_analysis} over 100 Monte Carlo trials at $L=750$ with $t\in[1000,2000]$. The black dashed line is a fit of the form $n(s)=As^{-\tau}$ with $\tau=2.24(10)$. (c) Determination of the fractal dimension, $d_f$. $s$ is the number of edges in each cluster and $R_g$ is its radius of gyration. Each data point is averaged over 100 Monte Carlo trials at $L=750$ for $t\in[1000,2000]$. The black dashed line is a fit of the form $\log R_g=\frac{1}{d_f}\log s + B$, yielding $d_f=1.8(2)$.
    }
    \label{fig:fig5}
\end{figure}

We now introduce a new model that retains the connectivity and re-routing constraints of the original model, while respecting the six-fold rotational spatial symmetry of the honeycomb lattice. Moreover, we implement the correction moves so that they never raise more flags than they lower, by considering two adjacent vertices rather than just one vertex. These two vertices are connected to 5 edges, yielding $2^5=32$ possible local flag configurations, which are all enumerated in Fig.~\ref{fig:fig4}. With this two-vertex configuration, up to two leaf moves can be applied (yielding both \textit{single leaf} and \textit{double leaf} moves), while the loop moves are applied irrespective of the purple and yellow triangular sublattices, and are only applied when $3$ flags can be lowered and at most $3$ flags will be raised.

\begin{figure}[!t]
    \centering
    \includegraphics[width=\columnwidth]{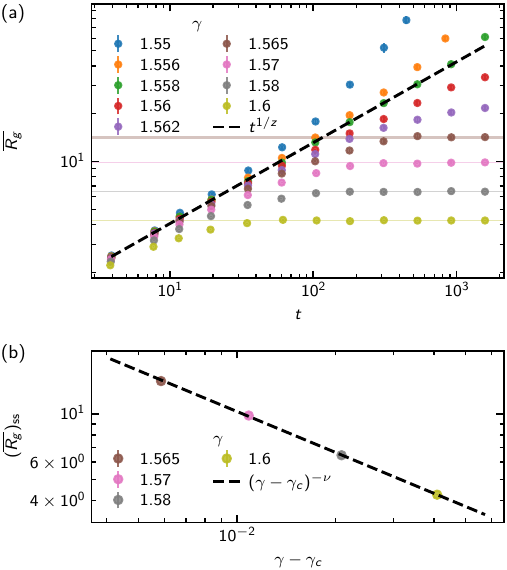}
    \caption{More isotropic model data: (a) The dynamic critical exponent, $z$. The mean radius of gyration over all clusters is defined as $\overline{R_g}=\sum_j (R_g)_j s_j^2/\sum_j s_j^2$ and computed for each cluster present at different times, $t$, over 100 Monte Carlo trials at $L=750$. Each data point corresponds to an average over a logarithmic time bin. The critical rate $\gamma_c$ is characterized by a power-law fit $\overline{R_g}=At^{1/z}$ at $\gamma=1.558$ with $z=1.96(10)$. The horizontal lines correspond to the steady-state mean radius of gyration for certain $\gamma>\gamma_c$, computed by averaging $\overline{R_g}$ over $t\in [1000,2000]$. (b) The steady-state mean radius of gyration, $(\overline{R_g})_{\text{ss}}$, as computed in (a) for $\gamma\ge 1.565$. The black dashed line is a fit of the form $(\overline{R_g})_{ss}=A(\gamma-\gamma_c)^{-\nu}$, yielding $\gamma_c=1.5591(5)$ and $\nu=0.63(2)$.}
    \label{fig:fig6}
\end{figure}

This isotropic model also has a hybrid percolation transition that appears to be of the same universality class as that of the previous anisotropic model.  In Fig.~\ref{fig:fig5}(a), the average flag density still jumps from around $\rho=0.2$ toward the absorbing state at $\rho=1$. This suggests that the Toom-like dynamics were not necessary for stabilizing the error channel in \cite{chirame2025stabilizing}.
We estimate that $\gamma_c=1.560(4)$ for this model, with fits of the critical exponents shown in Fig.~\ref{fig:fig5} and Fig.~\ref{fig:fig6} and more details available in Appendix~\ref{sec:appendix_convergence}. 

\begin{figure}[!t]
    \centering
    \includegraphics[width=\columnwidth]{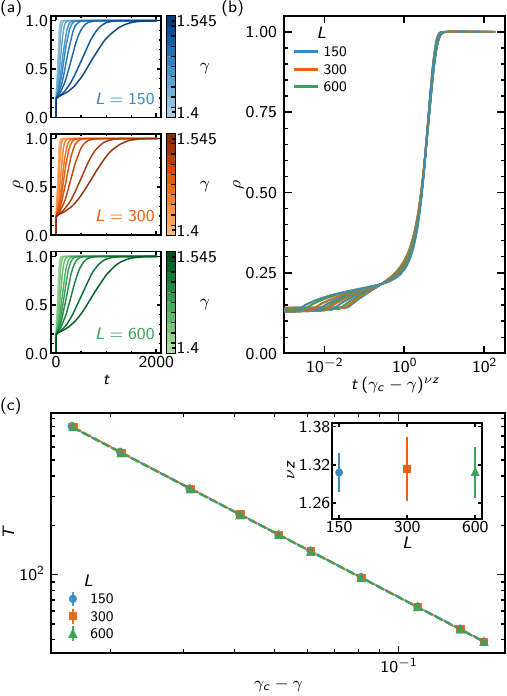}
    \caption{For the isotropic model: (a) Raw data of the mean density, $\rho$, over time for values of $\gamma\in[1.4,1.545]$ ($\gamma<\gamma_c$), initialized with $\rho=0$. (b) Scaling collapse of the mean density, based on the power-law fit in (c). The three colors correspond to different values of $L$. (c) Power-law fit of the time ($T$) for the mean density to cross a threshold $\rho_0=0.65$ versus $\gamma_c-\gamma$. Error bars are determined from varying $\rho_0\in[0.5,0.8]$. The power law exponent is $-\nu z$, with fitted values of $\nu z=1.31(3)$, $1.31(5)$, and $1.31(4)$ for $L=150,300,$ and $600$, while $\gamma_c=1.5610(10)$ for each $L$. All data are averaged over 100 Monte Carlo trials.}
    \label{fig:fig7}
\end{figure}

One further question is whether the phase transition from the active to absorbing states is caused by the same mechanism underlying the second-order percolation-type transition, or whether they just happen to coincide near each other. To test this, we look at how the mean flag density, $\rho$, increases over time on the absorbing side of the transition ($\gamma<\gamma_c$). In nucleation-driven first-order transitions, the characteristic time for transitions from the metastable active state to absorbing state is given by
\begin{equation}
T=(\gamma_c-\gamma)^{-2/3}\exp[c_{\text{nuc}}/(3(\gamma_c-\gamma))]\label{eq:activation}
\end{equation}
where $c_{\text{nuc}}$ is a parameter related to the line tension between the active state and absorbing state \cite{guo2007gtpc}. In Fig.~\ref{fig:fig7}(a) we show that the mean traces of the flag density over time for $\gamma\in[1.4,1.545]$ are $L$-independent for $L\ge 150$, which allows us to extract a characteristic timescale, $T$ as a function of $\gamma$ by finding where $\rho$ crosses certain thresholds $\rho_0$. We observe that $T$ varies as a power-law in $(\gamma_c-\gamma)$, rather than following the activation-barrier form of Eq.~\ref{eq:activation}. This suggests critical, barrier-free behavior, rather than a nucleation-driven transition. Moreover, if the same criticality driving the diverging correlation length for $\gamma>\gamma_c$ is responsible for this power-law decay of the characteristic timescale for reaching the absorbing state for $\gamma<\gamma_c$, then the characteristic timescale should scale as $T\sim (\gamma_c-\gamma)^{-\nu z}$, with $\nu$ and $z$ the correlation length exponent and dynamic critical exponent found previously. The fitted exponent of the power-law is $\nu z=1.31(5)$ across $L=150,300,600$ (Fig.~\ref{fig:fig7}(c)), which is consistent with the product $\nu\times z=0.63\times 1.96=1.23(7)$. This serves as evidence for the picture that the criticality at $\gamma_c$ via diverging cluster sizes seeds the first-order transition: below $\gamma_c$ the diverging clusters proliferate and merge without a barrier, with the time to fill in the lattice inheriting the critical timescale $T\sim(\gamma_c-\gamma)^{-\nu z}$. The final density is $\rho=1$ for all such $\gamma<\gamma_c$ since the fully flagged state is an absorbing state.

We note that the scaling collapse shown in Fig.~\ref{fig:fig7}(b) is that of the first-order transition where $\rho$ jumps from near $\rho_c$ to near $\rho=1$.  This plot results in a crossing of the curves near $\rho=\rho_c$.  The data near that crossing should instead be collapsed with critical scaling ($(\rho-\rho_c)(\gamma_c-\gamma)^{(\alpha-1)}$ on the vertical axis).  But such a critical scaling collapse requires the additional parameters $\rho_c$ and $\alpha$, and is found to have too many parameters to result in reliable estimates of those parameters.

\section{Removing the Absorbing State}
\label{sec:remove_absorb}

\begin{figure}[!t]
    \centering
    \includegraphics[width=\columnwidth]{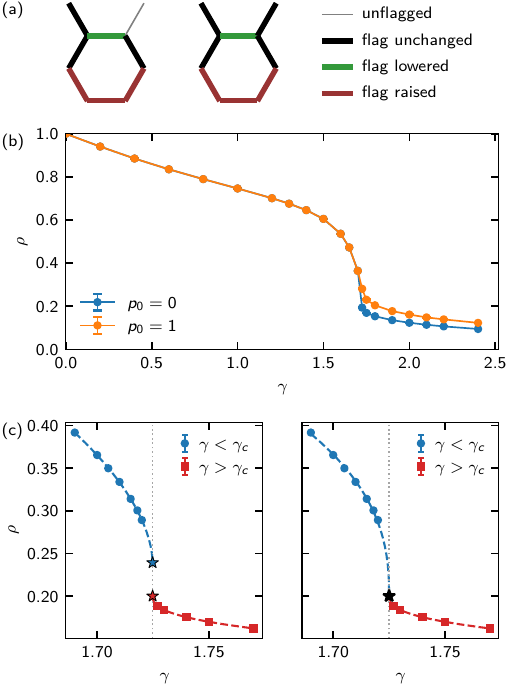}
    \caption{For the nonabsorbing model:  (a) Additional moves in the model for removing the absorbing state. These moves are applied with a probability $p_{\text{flip}}$ when the flags are in either of the configurations depicted. When all five edges are flagged, the flag on the middle edge is lowered, one of the two plaquettes bordering this edge is chosen at random, and flags are raised on all of the other edges bordering that plaquette (to preserve connections). 
    (b) The median flag density, $\rho$, for $p_{\text{flip}}=0.3$ over 100 Monte Carlo trials at $L=150$ with $t\in[1500,2000]$. The two different colors correspond to different initial states, where $p_0$ is the probability of a flag initially being raised.
    (c) The median flag density, $\rho$, for $p_{\text{flip}}=0.3$ over 100 Monte Carlo trials at $L=750$. Data from trials with $1.715\leq \gamma\leq 1.72$ are averaged over $t\in[9000,10000]$ while data at other $\gamma$ are averaged over $t\in[1800,2000]$. The blue circles ($\gamma<\gamma_c$) are initialized with $p_0=1$ and the red squares ($\gamma>\gamma_c$) are initialized with $p_0=0$. (Left) The dashed lines are fits $\rho_{\pm}=\rho_{c,\pm}+A_{\pm}(\gamma-\gamma_c)^{1-\alpha_{\pm}}$ where $\gamma_c$ is jointly fitted to both branches and $\rho_{c,\pm}$, $A_{\pm}$, and $\alpha_{\pm}$ are independently fit for each branch (with $M_+$ denoting the values for $\gamma>\gamma_c$). The fits yield $\alpha_+=0.61(3)$ and $\alpha_-=0.44(2)$. (Right) The joint fit is $\rho_{\pm}=\rho_{c}+A_{\pm}(\gamma-\gamma_c)^{\alpha_{\pm}}$, where $\gamma_c$ and $\rho_c$ are both jointly fitted and $A_{\pm}$ and $\alpha_{\pm}$ are independent for each branch. This yields $\alpha_+=0.62(2)$ and $\alpha_-=0.59(2)$.
    }
    \label{fig:fig8}
\end{figure}

The two models studied thus far, which exhibit a phase transition with both first- and second-order characteristics, have absorbing states. These absorbing states can be removed by allowing for additional isotropic local stochastic moves, which still forbid removing connections that could fragment a cluster. These moves, illustrated in Fig.~\ref{fig:fig8}(a), are performed in addition to all of the moves in Fig.~\ref{fig:fig4}, and allow for the lowering of flags in fully flagged configurations, thus removing the absorbing state. We generalize this model by only performing these additional moves with a probability $p_{\text{flip}}$ when such flag configurations are found, yielding an effective rate of $p_{\text{flip}}\gamma$. Therefore, this model is a generalization of the model explored in the previous section, which was the case $p_{\text{flip}}=0$. Over the rest of this section, we investigate the case of $p_{\text{flip}}=0.3$. We label this third model ``nonabsorbing''.

\begin{figure*}
    \centering
    \includegraphics[width=\textwidth]{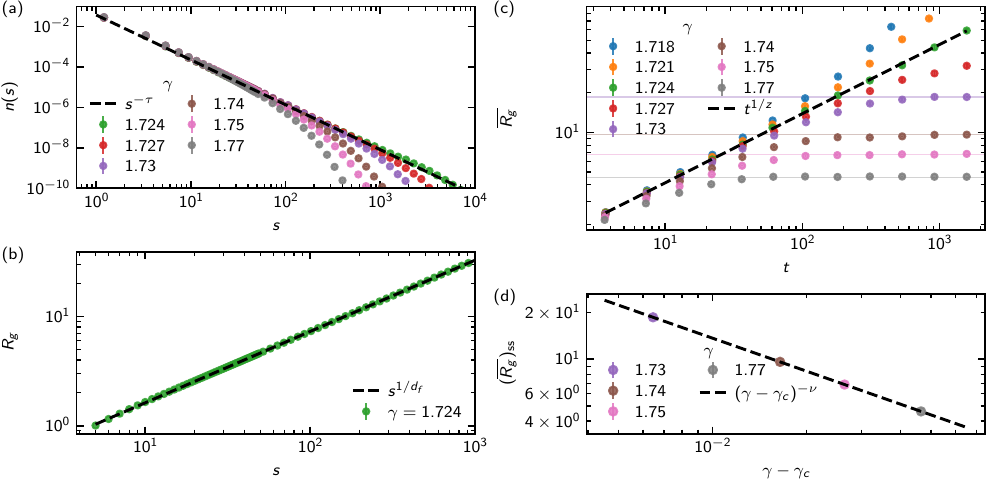}
    \caption{More nonabsorbing model data: (a) The frequency of having a cluster of size $s$, $n(s)$, near the transition at $\gamma_c=1.724(3)$. The black dashed line is a fit of the form $n(s)=As^{-\tau}$ with $\tau=2.22(5)$. (b) The radius of gyration, $R_g$, versus cluster size, $s$. The black dashed line is a fit of the form $\log R_g=\frac{1}{d_f}\log s + B$, yielding $d_f=1.53(5)$. (c) The average radius of gyration, $\overline{R_g}=\sum_j (R_g)_js_j^2/\sum_j s_j^2$, over time. The fitted dynamic critical exponent is $z=1.90(10)$. The horizontal lines correspond to the steady-state mean radius of gyration for certain $\gamma>\gamma_c$, computed by averaging $\overline{R_g}$ over $t\in[1000,2000]$.(d) The steady-state mean radius of gyration, as computed in (c) for $\gamma\ge 1.73$. The fitted correlation length exponent is $\nu=0.71(3)$ and fitted critical point is $\gamma_c=1.7235(5)$. (a)-(d) All data are averaged over 100 Monte Carlo trials at $L=750$, with data in (a) and (b) averaged over $t\in[1000,2000]$.}
    \label{fig:fig9}
\end{figure*}

The transition for this model with $p_{\text{flip}}=0.3$ appears to be either continuous or perhaps very weakly first-order.  The steady-state flag density for $\gamma$ just above the transition is still around $\rho\approx 0.2$, with a steep increase and possible small discontinuity near a critical $\gamma_c$. However, rather than jumping up to an absorbing state of $\rho=1$, the density increases from $\rho\approx 0.3$ to $\rho\approx 0.6$ over a small region $1.6\leq\gamma\leq1.72$, following an apparent power law (Fig.~\ref{fig:fig8}), and then smoothly rises toward $\rho=1$ as $\gamma\to 0$. We verify this by starting the system in the fully flagged state, which was previously an absorbing state, finding that the density converges to the same steady-state value for this initial condition as well as for an empty lattice for $\gamma<\gamma_c$. For $\gamma>\gamma_c$ (above the percolation transition), when starting the system in the fully flagged state, it is only possible to contract the initial infinite cluster into two intersecting noncontractible loops around the PBCs of the lattice (since connections still cannot be broken), resulting in a slight density offset between the two initial configurations in the low-density phase. However, this is a finite-size effect that should vanish in the limit $L\to \infty$ as the density of the two intersecting noncontractible loops goes to 0.

As before, we still find that $\frac{d\rho}{d\gamma}$ appears to diverge at $\gamma_c$, even though there is no longer a large jump in the average flag density. Using a system size of $L=750$, we fit exponents for $(\rho_c-\rho)\sim (\gamma-\gamma_c)^{1-\alpha}$ for both $\gamma\to\gamma_c^+$ ($\alpha_+$) and $\gamma\to\gamma_c^-$ ($\alpha_-$). When allowing $\rho_c$ to be independent between the two branches, it is slightly discontinuous, with $\rho_{c,+}=0.200(3)$ and $\rho_{c,-}=0.240(10)$, yielding fits of $\alpha_+=0.61(3)$ and $\alpha_-=0.44(2)$. However, a joint fit where $\rho_c$ is fixed to the same value in both limits, thereby making $\rho$ continuous across the percolation transition, is also consistent with the data (Fig.~\ref{fig:fig8}(c)), yielding $\alpha_+=0.62(2)$ and $\alpha_-=0.59(2)$. Distinguishing between these two cases would require simulations with $\gamma$ closer to $\gamma_c$, where the timescale for the system to reach the steady state diverges. Nevertheless, while we cannot conclude whether the transition is continuous with the present data, it is clear that the large jump in the flag density below the percolation threshold has been removed.

We confirm that for $p_{\text{flip}}=0.3$, the second-order features of the transition remain, and the critical exponents are not substantially changed from the other two models that instead have a large discontinuity in the density.  In Fig.~\ref{fig:fig9}, we identify that the percolation transition has shifted to a slightly higher critical point of $\gamma_c=1.724(3)$, coinciding with where $d\rho/d\gamma$ diverges, but that the critical exponents $\tau$, $d_f$, $z$, and $\nu$ are not significantly changed from the previously found values at $p_{\text{flip}}=0$. Therefore, all three models studied here appear to be in the same universality class in their critical behavior as $\gamma_c$ is approached from $\gamma>\gamma_c$.

\section{Discussion}

We have studied a class of active lattice percolation models, where the configurations are generated by a local stochastic process that adds, removes, and re-routes connections.  The removal of any connection that can possibly result in a single connected cluster being broken into more than one cluster is forbidden.  Such a connection may instead be re-routed in a way that maintains connections.
These models are inspired by error-correcting moves and a phase transition identified by Chirame, \textit{et al.}, who identified first-order characteristics of the phase transition~\cite{chirame2025stabilizing}. First, we showed that this phase transition is not purely first-order, but rather falls under the classification of hybrid percolation transitions, with diverging critical clusters at the transition point. The critical exponents differ from standard Bernoulli percolation, suggesting that the model belongs to a separate universality class.

Next, we showed that removing the spatial anisotropy of this original model and its Toom-like dynamics to make the model instead spatially isotropic with respect to the underlying honeycomb lattice preserved the hybrid percolation transition, with critical exponents that remained consistent with those of the original Toom-inspired model. 

Finally, we showed that removing the absorbing state from the newer isotropic model by allowing for additional moves (with a probability $p_{\text{flip}}=0.3$), that still never fragment clusters, yielded an apparently continuous percolation transition with still no substantial change in the critical exponents.  Further investigations may reveal if the absorbing state present in the first two models is essential to the discontinuous order parameter and hybrid percolation nature of their percolation transitions, or if the first-order and absorbing state aspects of these transitions can be made independent.
In Appendix~\ref{sec:appendix_pflip}, we show that at $p_{\text{flip}}=0.025$, the transition is still nearly continuous, albeit with a steeper increase in the mean flag density, $\rho$, near $\gamma_c$. For smaller values of $p_{\text{flip}}$, the autocorrelation time for the system to switch between low and high flag density phases near $\gamma_c$ reaches simulation timescales that are presently inaccessible.

In the Appendix we present additional results on finite-size and finite-time scaling and the dynamics of large compact clusters. 

Other possible future research directions include adapting these models to other types of lattices, allowing for the breaking of connectivity to study if never breaking connectivity is indeed an essential ingredient of this class of models, and applying analytical methods to these models.

\acknowledgements

We thank Fiona Burnell, Sanket Chirame, Sarang Gopalakrishnan, Nicholas O'Dea and Grace Sommers for discussions.  D.A.H. was supported in part by NSF QLCI grant OMA-2120757. The simulations were performed using the Adroit cluster managed by Princeton University Research Computing.

\section*{Appendix A: Methods and Data Analysis}

\setcounter{subsection}{0}
\renewcommand{\thesubsection}{\arabic{subsection}}

\makeatletter
\renewcommand{\p@subsection}{A}
\makeatother

\renewcommand{\theHsubsection}{appendixA.\arabic{subsection}}

\subsection{Monte Carlo Simulations}
\label{sec:appendix_monte_carlo}

We perform Monte Carlo simulations using asynchronous updates, following the loop logic of Chirame \textit{et al.} \cite{chirame2025stabilizing}. In particular, for the Toom-inspired model, the simulation works as follows.

\begin{algorithmic}
\State $\delta t = c_0 \coloneqq \frac{1}{\eta + 2\gamma/3}$
\While{$t \leq t_f$}
    \State save $\rho$ (average flag density)
    \If{$t = 10N\delta t$ for $N \in \mathbb{Z}$}
        \State save $\lvert x \rangle$ (state of all flags)
    \EndIf
    \For{$i = 1, 2, \ldots, L^2$}
        \State sample a random number $p \in [0,1]$ to choose the update rule
        \If{$p < c_0\eta$}
            \State sample a random integer $0 \leq j < L^2$ and raise the flag on edge $j$
        \Else
            \State sample a random integer $0 \leq v < 2L^2/3$ and apply the
                   correction operator at vertex $v$ (if possible)
        \EndIf
    \EndFor
    \State $t \to t + \delta t$
\EndWhile
\end{algorithmic}

For the isotropic model (and the generalized version of the isotropic model with $p_{\text{flip}}$), the constant $\delta t=c_0\coloneqq \frac{1}{\eta +2\gamma/3}$ is replaced by $\delta t=c_0\coloneqq \frac{1}{\eta +\gamma}$, while in the loop if $p\ge c_0 \eta$ then a random integer $0\leq j < L^2$ is picked, representing an edge, rather than a vertex, for the correction move. This is due to the fact that the correction moves are applied at ensembles of five edges, connected by two vertices, which can be indexed based on the central edge. The original factor of $2/3$ in the model from Chirame \textit{et al.} was in order to normalize the correction move rate $\gamma$ to be a \textit{per-edge} rate, hence in the isotropic model this factor is no longer necessary.

\subsection{Cluster and critical exponent analysis}
\label{sec:appendix_cluster_analysis}

In the Monte Carlo loop described above, we save the state of all flags on the lattice every $10$ increments of $\delta t$. We post-process this flag data to find all clusters using a breadth-first search. We track the net displacement of each extra edge found during this search in order to reconstruct each cluster as a connected component even if it spans across the periodic boundary conditions on the original lattice, allowing for calculations of the radius of gyration. Furthermore, whenever an edge is revisited during the BFS, we check if the net displacement was non-zero, which would imply that the cluster forms a non-contractible loop around the periodic boundary conditions of the lattice, thereby forming an infinite length cluster. For all computations of critical exponents, we ensured that the lattice size, $L$, and final time used, $t_f$, were chosen such that there are no such clusters across all Monte Carlo trials.

We also now describe how we process the raw cluster statistics found in each Monte Carlo trial. Generally, since data within each trial may be highly correlated, we first average data within each trial, prior to averaging across all trials with equal weighting. Error bars are calculated via the standard error on the mean from the residuals of the trial means about it. 

For example, for the Fisher exponent, $\tau$, we first average $n(s)$ within each trial. We average it across all clusters found in a given time window ($t\in [1000,2000]$ for figures in the main text), at all integer cluster sizes $s\leq 50$, as well as within logarithmically scaled bins for $s>50$. Then, the final $n(s)$ presented in a given figure is computed by averaging the $n(s)$ from each of 100 Monte Carlo trials, with error bars as above. We similarly average $R_g$ individually over cluster sizes $s\leq 50$ or over logarithmically scaled bins for $s>50$, first within trials.

For the dynamic critical exponent, $z$, we compute $\overline{R_g}$ across all clusters found in logarithmically scaled time intervals, first within each trial prior to averaging over all trials. For the correlation length exponent, $\nu$, we average $\overline{R_g}$ over a single time interval $t\in[1000,2000]$ within each trial, before averaging over all trials.

\subsection{Convergence of cluster exponents with respect to $L$ and $t_f$}
\label{sec:appendix_convergence}

\begin{figure*}[!t]
    \centering
    \includegraphics[width=\textwidth]{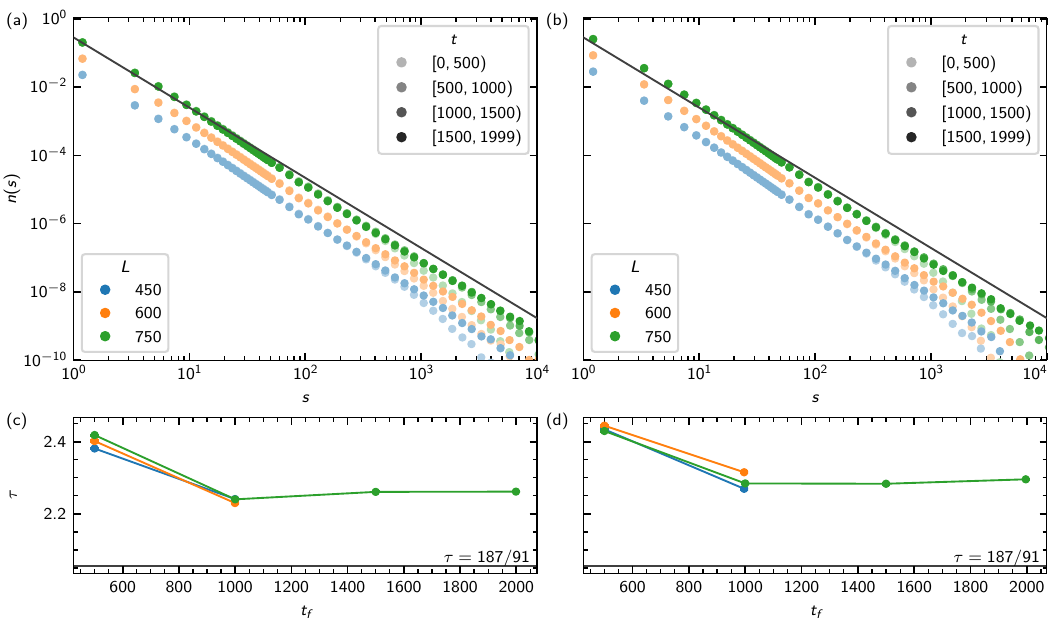}
    \caption{Convergence of the estimate for the Fisher exponent, $\tau$, for the Toom-inspired
    and isotropic models with respect to $L$ and $t$. (a,b) $n(s)$ versus $s$ for various system sizes, $L$ (different colors), and time-averaging windows (different shades). The points for different system sizes are scaled by an offset for clearer visibility. The solid black line is of the form $n(s)=As^{-187/91}$, which is the Fisher exponent for 2D Bernoulli percolation. (a) The Toom-inspired model with $\gamma=6.455$. (b) The isotropic model with $\gamma=1.56$. (c,d) The fitted Fisher exponent, $\tau$, for each $L$ and each time window $t\in[t_f-500,t_f)$. $\tau$ is fitted in the range $s\in[10^2,10^3]$, with error bars denoting the statistical error from the fit. By varying the fitting window and using data for other $\gamma$, we estimate $\tau=2.26(10)$ for the Toom-inspired model and $\tau=2.24(10)$ for the isotropic model.}
    \label{fig:fig10}
\end{figure*}

We demonstrate that the geometric critical exponents, $\tau$ and $d_f$, are well-converged with respect to the lattice size and timescale, for both the Toom-inspired and isotropic models. We study lattice sizes $L=450$, $600$, and $750$, with simulations up to $t_f=1000, 1000,$ and $2000$. In particular, we show in Fig.~\ref{fig:fig10} that $n(s)$ is well-converged across all three of these system sizes for $t\ge 500$ up to $s=10^3$. We correspondingly fit $\tau$ for $s\in[10^2,10^3]$, so that the fit is not dominated by small clusters or larger clusters where $n(s)$ has not converged. This yields small statistical uncertainties, so we approximate the uncertainty on $\tau$ by comparing fitted values across different $\gamma$ near $\gamma_c$, and by extending the fitting range.
We also present the data with a comparison to the ordinary percolation value of $\tau=\frac{187}{91}$, finding that $\tau$ for the models we study is visibly larger.

\begin{figure*}[!t]
    \centering
    \includegraphics[width=\textwidth]{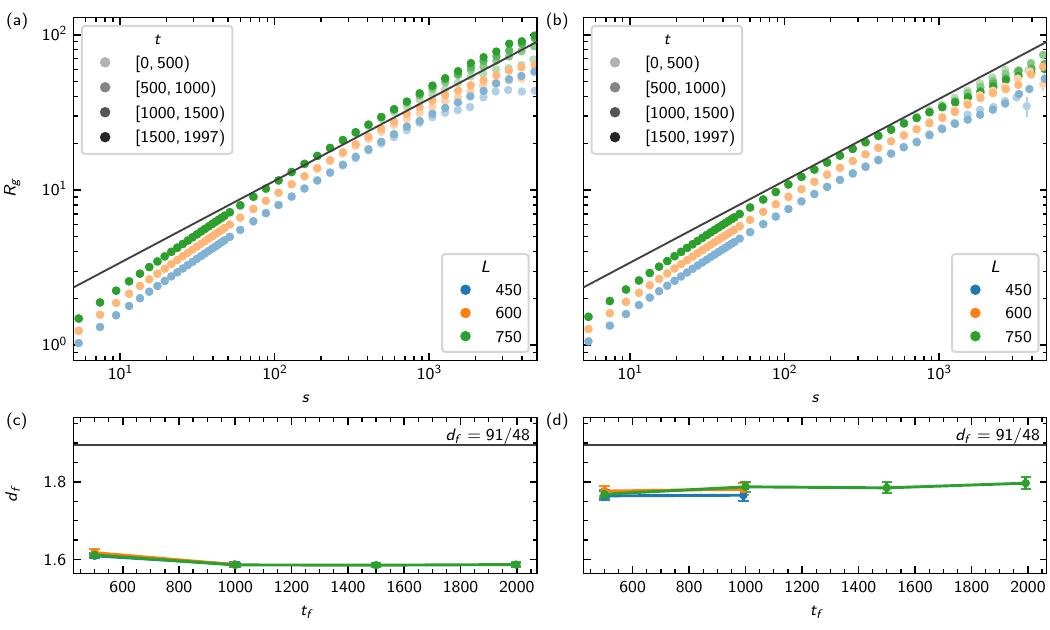}
    \caption{Convergence of the estimate for the fractal dimension, $d_f$, for the Toom-inspired and isotropic
    models with respect to $L$ and $t$. (a,b) The radius of gyration $R_g$ versus $s$ for various system sizes, $L$ (different colors), and time-averaging windows (different shades). The points for different system sizes are scaled by an offset for clearer visibility. The solid black line is of the form $\log R_g=\frac{48}{91}\log s + B$, where $d_f=\frac{91}{48}$ is the fractal dimension for 2D Bernoulli percolation. (a) The Toom-inspired model with $\gamma=6.455$. (b) The isotropic model with $\gamma=1.56$. (c,d) The fitted fractal dimension, $d_f$, for each $L$ and each time window $t\in[t_f-500,t_f)$. The fractal dimension, $d_f$, is fitted in the range $s\in[10^2,10^3]$, with error bars denoting the statistical error from the fit. By varying the fitting window and using data for other $\gamma$, we estimate $d_f=1.59(8)$ for the Toom-inspired model and $d_f=1.8(2)$ for the isotropic model.}
    \label{fig:fig11}
\end{figure*}

We study the fractal dimension, $d_f$, for both models over the same simulations and similar cluster masses (Fig.~\ref{fig:fig11}). Interestingly, we do find differences between the two models, where within fixed $s$-bins, $R_g$ appears to increase over time for the Toom-inspired model, while decreasing over time for the isotropic model. In the Toom-inspired model, this means that at large cluster masses, $R_g$ is below the power law trend at shorter simulation timescales, but by averaging over later windows it appears to approach the same power-law up to larger cluster masses. In particular, we fit a power-law for $s\in[10^2,10^3]$ with small statistical uncertainty, and the trend towards larger $s$ appears to indicate that this power-law fit will continue provided longer simulation timescales. Contrastingly, for the isotropic model, $R_g$ continues to decrease for larger cluster masses. This indicates that the larger clusters are more filled-in over time, thus, it is possible that the value of $\gamma=1.56$ may be slightly below $\gamma_c$. However, data for higher $\gamma$ (i.e. $\gamma=1.565$) does not have an appreciable number of clusters of large mass ($s\ge 10^3$) since $n(s)$ drops off from the power-law scaling, making it difficult to observe what happens at large $s$. For $\gamma=1.56$, we still fit $d_f$ in the range $s\in[10^2,10^3]$, finding $d_f=1.80\pm 0.02_{\text{stat}}$, but estimate a larger uncertainty of $d_f=1.8(2)$ since adjusting the cluster sizes used in the fitting window yields a large change in $d_f$. Thus, the fractal dimension in the isotropic model may be consistent with the fractal dimension for Bernoulli percolation, $d_f=\frac{91}{48}\approx 1.896$. Further investigation is also necessary to determine whether this contrast in how the largest clusters evolve across the two models means that they do not share a universality class, or whether it is a result of some finite-size or off-critical effects.

\begin{figure}[!t]
    \centering
    \includegraphics[width=\columnwidth]{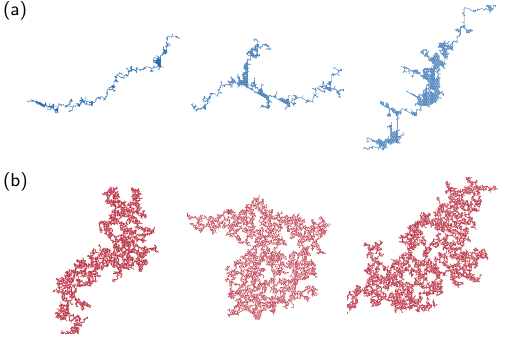}
    \caption{(a) Largest cluster across three separate trials at $t=2000$ for the Toom-inspired model, with $\gamma=6.455$ and $L=750$. (b) Largest cluster for Bernoulli bond percolation on a honeycomb lattice at $p=0.6327<p_c=0.6527$ and $L=750$. This value of $p$ is chosen such that the clusters have a similar correlation length as the clusters in (a).}
    \label{fig:fig12}
\end{figure}

We also present the largest clusters at $t_f=2000$ across three different trials at $L=750$ and $\gamma=6.455$ for the Toom-inspired model in Fig.~\ref{fig:fig12}, with a comparison to Bernoulli percolation clusters. Since the radius of gyration is finite at this fixed $t$ for the Toom-inspired model, we use $p=0.6327<p_c=0.6527$ for the Bernoulli percolation clusters in order to obtain clusters of similar length scales. This further demonstrates the qualitative difference between the clusters we study and Bernoulli percolation clusters, as the latter are more space-filling due to their higher fractal dimension.

\section*{Appendix B: Dynamics of large droplets}

\makeatletter
\renewcommand{\p@subsection}{B}
\makeatother

\renewcommand{\theHsubsection}{appendixB.\arabic{subsection}}

\begin{figure}[!t]
    \centering
    \includegraphics[width=\columnwidth]{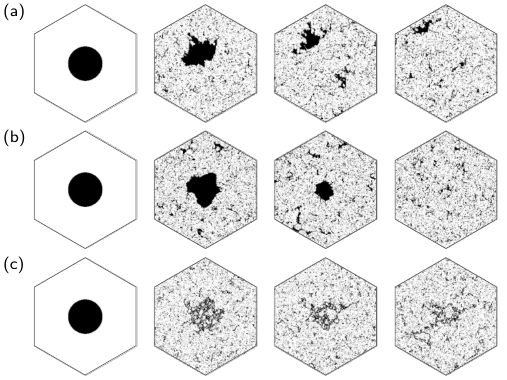}
    \caption{Evolution of a circular region of flags under each model near the critical point, on a lattice with $L=300$ and a circular region of radius $r=50$. (a) The model from Chirame \textit{et al.} with $\gamma=6.455$. The time interval between panels is $\Delta t=50$. (b) The isotropic model with $\gamma=1.562$ and $\Delta t=300$. (c) The nonabsorbing model with $p_{\text{flip}}=0.3$, with $\gamma=1.724$ and $\Delta t=50$.}
    \label{fig:fig13}
\end{figure}

As shown in Sec.~\ref{sec:six_fold}, the scaling of the time it takes for the system to reach the absorbing state in the isotropic model, on the absorbing side of the transition ($\gamma<\gamma_c$), is consistent with critical, barrier-free behavior driven by the same critical dynamics on the active side of the transition ($\gamma>\gamma_c$). This serves as evidence that the discontinuous order parameter, and first-order nature of the transition, do not arise from nucleation-mediated dynamics. We further study nucleation by initializing the system in a state with a large, circular domain of flags (Fig.~\ref{fig:fig13}).

We find that at $\gamma\gtrsim\gamma_c$, these large droplets can be erased in all three of the models investigated in this paper. In the Toom-inspired model, this circular domain is erased starting from the SE corner and moves towards the NW, due to the spatial anisotropy of the loop moves. In the other two models (the isotropic model with $p_{\text{flip}}=0$ and $p_{\text{flip}}=0.3$), the circular domain is compressed from all directions. At $p_{\text{flip}}=0.3$, gaps appear within the domain, due to the absence of an absorbing state.

\section*{Appendix C: Results with $p_{\text{flip}}=0.025$}
\phantomsection

\makeatletter
\def\@currentlabel{C}
\makeatother
\label{sec:appendix_pflip}

\makeatletter
\renewcommand{\p@subsection}{C}
\makeatother

\renewcommand{\theHsubsection}{appendixC.\arabic{subsection}}

We showed in Sec.~\ref{sec:remove_absorb} that by adding additional moves at a rate of $p_{\text{flip}}=0.3$, the order parameter ($\rho$) appeared continuous at a system size of $L=150$. We similarly show in this section that it is nearly continuous at $p_{\text{flip}}=0.025$, albeit with apparent bistability at smaller system sizes that vanishes with larger $L$.

We study system sizes between $L=15$ and $L=90$ in order to run trials longer than the autocorrelation time of the density. We define the autocorrelation of the density as
\begin{equation}
G(t)=\langle \rho(t_0)\rho(t_0+t)\rangle_{t_0}-\langle \rho(t_0)\rangle_{t_0}^2
\end{equation}
where the average is taken over $t_0 \ge T_{\text{MIN}}$. If the system exhibits apparent bistability, with a characteristic time for the state of the system to switch between two bistable phases, then $G(t)$ should decay on the order of the characteristic switching time. By running individual trials for significantly longer than this switching time, we can ensure that the probability density of the mean density of the system is well-sampled. For example, at $L=15$, across $100$ Monte Carlo trials we fit the average decay time constant of $G(t)$ to be $\tau_{1/e}=350(30)$, so by running trials out to $t_f=10^5$, the system will have not only forgotten its initial state, but also sampled over both bistable phases, as well as the valley between them. 

\begin{figure}[!t]
    \centering
    \includegraphics[width=\columnwidth]{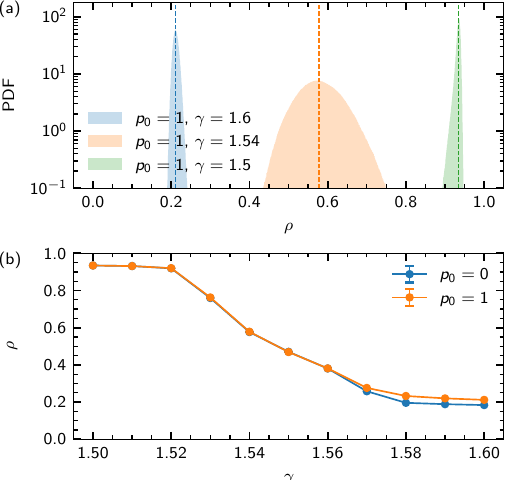}
    \caption{(a) Histograms of the steady-state flag density, $\rho$, for this model with $p_{\text{flip}}=0.025$. The data are averaged over 100 Monte Carlo trials at $L=150$ with $t\in[10000,50000]$, starting from a fully flagged lattice. (b) The median flag density, $\rho$, for $p_{\text{flip}}=0.025$ over 100 Monte Carlo trials at $L=150$ with $t\in[10000,50000]$. The two different colors correspond to different initial states, where $p_0$ is the probability of a flag initially being raised.}
    \label{fig:fig14}
\end{figure}

\begin{figure*}[!b]
    \centering
    \includegraphics[width=\textwidth]{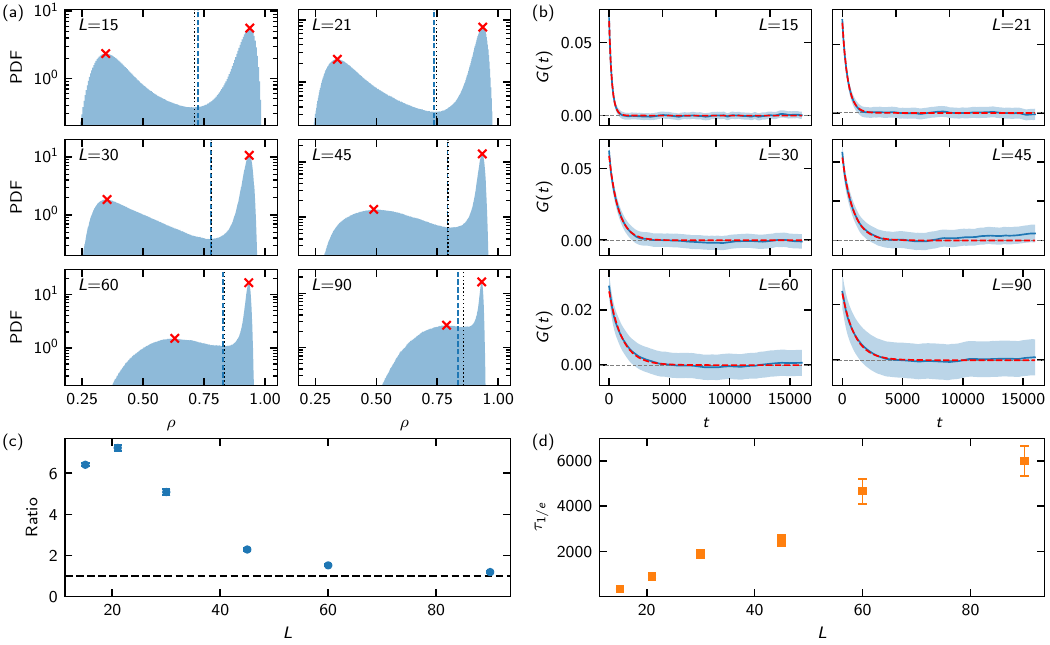}
    \caption{(a) Probability density of the average flag density at different lattice sizes. The correction rate, $\gamma$, is chosen such that the minimum of the probability density function (black vertical dotted line) coincides with the median average flag density (blue vertical dashed line). The red crosses mark the maxima of two peaks. Data are averaged over $t\in[10^4,10^5]$ for $L\leq 45$ and $t\in[2\times 10^4,10^5]$ for $L\ge 60$. (b) Temporal autocorrelation $G(t)=\langle \rho(t_0)\rho(t_0+t)\rangle -\langle \rho \rangle ^2$ where the expectation value is taken over $t_0 \ge T$ (with $T$ determined based on the autocorrelation decay time). We use $T=10^4$ for $L\leq 45$ and $T=2\times10^4$ for $L\ge 60$. The red dashed line is an exponential fit. (c) The ratio of the probability density function (PDF) at the smaller of the two maxima (red cross, left peak) to the PDF at its minimum (black line) for each of the panels in (a). Error bars are determined by finding the ratio over 100 Monte Carlo trials. (d) Decay of the autocorrelation as fitted in (b). Error bars are determined by data from 100 Monte Carlo trials.} 
    \label{fig:fig15}
\end{figure*}

To examine how the bistability disappears, we calibrate $\gamma$ so that the median of $\rho$ coincides with the minimum of the probability density of $\rho$ at each system size, $L$. We identify the locations of two peaks in the probability density of $\rho$ as well as the minimum by looking at the average probability density across 100 Monte Carlo trials, and then fit the ratio of the depth barrier per trial, by reading off the probability density function (PDF) at the left (shorter) peak and at its minimum. We average this ratio across all trials, and find that it decreases with $L$, nearly approaching $1$ by $L=90$ (Fig.~\ref{fig:fig15}(c)). This is in contrast to a first-order transition, where the depth ratio corresponds to a free energy barrier that would increase with the system size. The left peak of the probability density also moves towards higher $\rho$ at the calibration $\gamma$ (which decreases with $L$), further serving as evidence that the bistability is a finite size effect and disappears as two peaks merge into one continuous, Gaussian-shaped peak at larger $L$ (Fig.~\ref{fig:fig14}(a)). The characteristic decay of the autocorrelation function increases with $L$, but stays below $10^4$, so that simulations can still be run out to $t_f=10^5$. At smaller $p_{\text{flip}}<0.025$, this autocorrelation decay timescale increases more rapidly with $L$, making the necessary simulation timescales higher than what is presently accessible.

At $p_{\text{flip}}=0.025$, for $L=150$, the transition appears nearly continuous, with a single peak in the histogram of the mean flag density, $\rho$, that moves across the range $0.2<\rho<0.9$ for $1.58>\gamma>1.52$ (see Fig.~\ref{fig:fig14}). As before, we see that whether starting the system in the empty ($p_0=0$) or fully flagged ($p_0=1$) state, the density converges to the same value for $t>10000$.

\clearpage

\bibliography{main}

\end{document}